\documentclass{ws-ijmpd}

\begin{document}

\markboth{Melia}
{Constraints on Dark Energy}

%
\catchline{}{}{}{}{}
%

\title{Constraints on Dark Energy from the Observed Expansion\\
of our Cosmic Horizon}

\author{Fulvio Melia}

\address{Department of Physics and Steward Observatory,\\
The University of Arizona,\\
Tucson, AZ 85721, USA\\
melia@physics.arizona.edu}

\maketitle

\begin{history}
\received{1 December 2008}
\accepted{27 December 2008}
\end{history}

\begin{abstract}
Within the context of standard cosmology, an accelerating universe requires
the presence of a third `dark' component of energy, beyond matter and radiation.
The available data, however, are still deemed insufficient to distinguish between
an evolving dark energy component and the simplest model of a time-independent
cosmological constant. In this paper, we examine the cosmological expansion
in terms of observer-dependent coordinates, in addition to the more conventional
co-moving coordinates. This procedure explicitly reveals the role played by the
radius $R_{\rm h}$ of our cosmic horizon in the interrogation of the data. (In
Rindler's notation, $R_{\rm h}$ coincides with the `event horizon' in the case
of de Sitter, but changes in time for other cosmologies that also contain matter
and/or radiation.) With this approach, we show that the interpretation of dark
energy as a cosmological constant is clearly disfavored by the observations.
Within the framework of standard Friedman-Robertson-Walker cosmology, we derive
an equation describing the evolution of $R_{\rm h}$, and solve it using the WMAP
and Type Ia supernova data. In particular, we consider the meaning of the
\emph{observed} equality (or near equality) $R_{\rm h}(t_0)\approx ct_0$, where
$t_0$ is the age of the Universe. This empirical result is far from trivial, for a
cosmological constant would drive $R_{\rm h}(t)$ towards $ct$ (where $t$ is the
cosmic time) only once---and that would have to occur right now. Though we are not
here espousing any particular alternative model of dark energy, for comparison we
also consider scenarios in which dark energy is given by scaling solutions, which
simultaneously eliminate several conundrums in the standard model, including the
`coincidence' and `flatness' problems, and account very well for the fact that
$R_{\rm h}(t_0)\approx ct_0$.
\end{abstract}


\section{Introduction}	

Over the past decade, Type Ia supernovae have been used successfully as standard
candles to facilitate the acquisition of several important cosmological parameters.
On the basis of this work, it is now widely believed that the Universe's expansion
is accelerating.$^{1,2}$
In standard cosmology, built on the assumption of spatial homogeneity and isotropy,
such an expansion requires the existence of a third form of energy, beyond the basic
admixture of (visible and dark) matter and radiation.

One may see this directly from the (cosmological) Friedman-Robertson-Walker (FRW)
differential equations of motion, usually written as
\begin{equation}
H^2\equiv\left(\frac{\dot a}{a}\right)^2=\frac{8\pi G}{3c^2}\rho-\frac{kc^2}{a^2}\;,
\end{equation}
\begin{equation}
\frac{\ddot a}{a}=-\frac{4\pi G}{3c^2}(\rho+3p)\;,
\end{equation}
\begin{equation}
\dot\rho=-3H(\rho+p)\;,
\end{equation}
in which an overdot denotes a derivative with respect to cosmic time $t$, and $\rho$
and $p$ represent, respectively, the total energy density and total pressure.
In these expressions, $a(t)$ is the expansion factor, and $(r,\theta,\phi)$ are
the coordinates in the comoving frame. The constant $k$ is $+1$ for a closed
universe, $0$ for a flat universe, and $-1$ for an open universe.

Following convention, we write the equation of state as $p=\omega\rho$.
A quick inspection of Eq.~(2) shows that an accelerated expansion
($\ddot a>0$) requires $\omega<-1/3$. Thus, neither radiation ($\rho_r$, with
$\omega_r=1/3$), nor (visible and dark) matter ($\rho_m$, with $\omega_m\approx 0$)
can satisfy this condition, leading to the supposition that a third `dark'
component $\rho_d$ (with $\omega_d<-1/3$) of the energy density $\rho$ must
be present. In principle, each of these contributions to $\rho$ evolves
according to its own dependence on $a(t)$.

Over the past few years, complementary measurements$^3$
of the cosmic microwave background (CMB) radiation have indicated that the Universe
is flat (i.e., $k=0$), so $\rho$ is at (or very near) the ``critical" density
$\rho_c\equiv 3c^2H^2/8\pi G$. But among the many peculiarities of the standard
model is the inference, based on current observations, that $\rho_d$ must
itself be of order $\rho_c$. Dark energy is often thought to be the manifestation
of a cosmological constant, $\Lambda$, though no reasonable explanation has yet been
offered as to why such a fixed, universal density ought to exist at this scale. It is
well known that if $\Lambda$ is associated with the energy of the vacuum in quantum
theory, it should have a scale representative of phase transitions in the early
Universe---many, many orders of magnitude larger than $\rho_c$.

Many authors have attempted to circumvent these difficulties by proposing
alternative forms of dark energy, including Quintessence,$^{4,5}$
which represents an evolving canonical scalar
field with an inflation-inducing potential, a Chameleon field$^{6,7,8}$
in which the
scalar field couples to the baryon energy density and varies from solar system to
cosmological scales, and modified gravity, arising out of both string motivated,
or General Relativity modified actions,$^{9,10,11}$ which
introduce large length scale corrections modifying the late time evolution
of the Universe. The actual number of suggested remedies is far greater
than this small, illustrative sample.

Nonetheless, though many in the cosmology community suspect that some sort
of dynamics is responsible for the appearance of dark energy, until now the
sensitivity of current observations has been deemed insufficient$^7$
to distinguish between an evolving dark energy component and the simplest
model of a time-independent cosmological constant $\Lambda$. This conclusion,
however, appears to be premature, given that constraints on the universe's
expansion arising from the observed behavior of our cosmic horizon has not
yet been fully folded into the interrogation of the current data. The purpose
of this paper is to demonstrate that a closer scrutiny of the available
measurements, if proven to be reliable, can in fact already delineate
between evolving and constant dark energy theories, and that a simple
cosmological constant $\Lambda$, characterized by a fixed $\omega_d\equiv
\omega_\Lambda=-1$, is disfavored by the observations.

\section{The Cosmic Horizon}

In an earlier paper,$^{13}$ we introduced a transformation of the
Robertson-Walker (RW) metric (from which Eqs.~1, 2, and 3 are derived)
into a new set of (observer-dependent) coordinates $(cT,R,\theta,\phi)$,
where $R\equiv a(t)r$ and $T(R)$ is the time (in the observer's frame)
corresponding to the radius $R$. The cosmic time $t$ coincides with $T$
only at the origin, i.e., $T(0)=t$. For all other radii, $T$ is dilated
relative to $t$ from the effects of curvature induced by the
mass-energy content of the universe. Ironically, de Sitter's
own metric was first written in terms of these observer-dependent
coordinates,$^{14}$ 
\begin{equation}
ds^2=\left[1-\left({R\over R_{\rm h}}\right)^2\right]c^2\,dT^2-
{dR^2\over 1-\left({R/R_{\rm h}}\right)^2}-R^2\,d\Omega^2\;,
\end{equation}
but this appears to have long since been forgotten, and everyone now uses 
the form of the metric written in terms of $r$ and $t$ only:
\begin{equation}
ds^2=c^2\,dt^2-a^2(t)[dr^2(1-kr^2)^{-1}+r^2(d\theta^2+\sin^2\theta\,d\phi^2)]\;.
\end{equation}

However, it is easy to see why it makes sense to consider both sets of coordinates,
because written in terms of $(cT,R,\theta,\phi)$, the RW metric explicitly
reveals the dependence of $dT$ and $dR$ on the characteristic radius $R_{\rm h}$,
defined by the condition
\begin{equation}
\frac{2GM(R_{\rm h})}{c^2}=R_{\rm h}\;.
\end{equation}
In this expression, $M(R_{\rm h})=(4\pi/3) R_{\rm h}^3\rho/c^2$ is the enclosed mass at
$R_{\rm h}$. In terms of $\rho$, we find that 
\begin{equation}
R_{\rm  h}=({3c^4/8\pi G\rho})^{1/2}
\end{equation}
or, more simply, $R_{\rm h}=c/H_0$ in a flat universe. Not surprisingly, this is
the radius at which a sphere encloses sufficient mass-energy to create an infinite
redshift (i.e., $T(R)\rightarrow\infty$ as $R\rightarrow R_{\rm h}$) as seen by an
observer at the origin of the coordinates $(cT,R,\theta,\phi)$.

When the Robertson-Walker metric is written in terms of $(cT,R,\theta,\phi)$, the
presence of $R_{\rm h}$ alters the intervals of time we measure (using the clocks
fixed to our origin) progressively more and more as $R\rightarrow R_{\rm h}$. And
since the gravitational time dilation becomes infinite at $R_{\rm h}$, it is
physically impossible for us to see any process occurring beyond this radius.
Light emitted beyond $R_{\rm h}$ is infinitely redshifted by the time it
reaches us and it therefore carries no signal.

This is precisely the reason why the recent observations have a profound impact
on our view of the cosmos. The Hubble Space Telescope Key Project on the extragalactic
distance scale has measured the Hubble constant $H$ with unprecedented accuracy,$^{15}$
yielding a current value $H_0\equiv H(t_0)=71\pm6$ km s$^{-1}$ Mpc$^{-1}$.
(For $H$ and $t$, we will use subscript ``0" to denote cosmological values pertaining
to the current epoch.) With this $H_0$, we infer that $\rho(t_0)=\rho_c\approx 
9\times 10^{-9}$ ergs cm$^{-3}$.

Given such precision, it is now possible to accurately calculate the radius $R_{\rm h}$.
From the Hubble measurement of $\rho(t_0)$, we infer that $R_{\rm h}\approx 13.5$ billion
light-years; this is the maximum distance out to which measurements of the cosmic
parameters may be made at the present time. At first glance, it may seem that it
had to be this way, since the age $t_0$ of the universe is also known to be $13.7$
billion years. But in fact, the FRW equations predict that $R_{\rm h}$ should
{\it not} be equal to $ct_0$, unless $\omega$ has a very special value.

Let us consider how the radius $R_{\rm h}$ evolves with the universal expansion.
Clearly, in a de Sitter universe$^{14}$ with a constant $\rho$ (proportional to $\Lambda$),
$R_{\rm h}$ is fixed forever.\footnote{In this situation, $R_{\rm h}$
coincides with the event horizon defined by Rindler.$^{16}$} But for any universe
with $\omega\not=-1$, $R_{\rm h}$ must be a function of time. From the definition
of $R_{\rm h}$ and Eqn.~(3), it is easy to see that
\begin{equation}
{\dot R}_{\rm h}=\frac{3}{2}(1+\omega)\,c\;,
\end{equation}
a remarkably simple expression that nonetheless leads to several important
conclusions regarding our cosmological measurements. We will use it here
to distinguish between constant and evolving dark energy theories. Notice,
for example, that ${\dot R}_{\rm h}=c$ only for the special case $\omega=-1/3$,
which is not consistent with the currently favored $\Lambda$CDM model of
cosmology.

Take $t$ to be some time in the distant past (so that $t\ll t_0$). Then,
integrating Eqn.~(8) from $t$ to $t_0$, we find that
\begin{equation}
R_{\rm h}(t_0)-R_{\rm h}(t)=\frac{3}{2}(1+\langle\omega\rangle)\,ct_0\;,
\end{equation}
where
\begin{equation}
\langle\omega\rangle\equiv \frac{1}{t_0}\int_{t}^{t_0}\omega\,dt
\end{equation}
is the time-averaged value of $\omega$ from $t$ to the present time.

Now, for any $\langle\omega\rangle>-1$, $\rho$ drops as the universe
expands (i.e., as $a(t)$ increases with time), and since
$R_{\rm h}\sim\rho^{-1/2}$, clearly $R_{\rm h}(t)\ll R_{\rm h}(t_0)$. Therefore,
\begin{equation}
R(t_0)\approx\frac{3}{2}(1+\langle\omega\rangle)\,ct_0\;.
\end{equation}
The reason we can use the behavior of $R_{\rm h}$ as the universe expands
to probe the nature of dark energy is that the latter directly impacts
the value of $\langle\omega\rangle$. A consideration of how the cosmic
horizon $R_{\rm h}$ evolves with time can therefore reveal whether or not
dark energy is dynamically generated. Indeed, we shall see shortly
that the current observations, together with Eqn.~(11), are
already quite sufficient for us to differentiate between the various models.

Before we do that, however, we can already see from this expression
that there may be a serious problem with our interpretation of $t_0$ in
the standard model of cosmology. From WMAP observations,$^3$
we infer that the age $t_0$ of the universe is $\approx 13.7$ billion years.
Since $R_{\rm h}\approx 13.5$ billion light-years, this can only occur if
$\langle\omega\rangle\le -1/3$. Of course, this means that the existence
of dark energy (with such an equation of state) is required by the
WMAP and Hubble observations alone, independently of the Type Ia
supernova data. But an analysis of the latter (see \S~4 below, particularly
Figs.~5 and 8) reveals that the value $\langle\omega\rangle=-1/3$ is 
apparently ruled out, so in fact $\langle\omega\rangle<-1/3$.

But this means that $R_{\rm h}\not=ct_0$; in fact, $R_{\rm h}$ must be
less than $ct_0$, which in turns suggests that the universe is older than
we think. Unless $\omega=-1/3$, what we infer to be the time since the Big
Bang, is instead the ``horizon" time $t_h\equiv R_{\rm h}/c$, which must be
shorter than $t_0$. This may seem absurd at first, but we must remember
that any events occurring beyond $R_{\rm h}$ are not visible to us yet (or
ever, depending on whether or not $R_{\rm h}$ is constant). In addition,
$T(R)$ becomes progressively more dilated as we view events taking place
closer and closer to $R_{\rm h}$, so that we see the universe as it
appeared just after the Big Bang when $R\approx R_{\rm h}$, regardless
of when the Big Bang actually occurred, as long as $t_0\ge R_{\rm h}/c$.

This phenomenon has some important consequences that may resolve
several long-standing conflicts in cosmology. Through our analysis below,
we will gain a better understanding of how it works, which will permit us
to calculate $t_0$ more precisely.

\section{The Cosmological Constant}

The standard model of cosmology contains a mixture of cold dark matter
and a cosmological constant with an energy density fixed at the
current value, $\rho_d(t)\equiv\rho_\Lambda(t)\approx 0.7\,\rho_c(t_0)$,
and an equation of state with $\omega_d\equiv\omega_\Lambda=-1$. Known
as $\Lambda$CDM, this model has been reasonably successful in
accounting for large scale structure, the cosmic microwave background
fluctuations, and several other observed cosmological
properties.$^{3,17,18}$
\begin{figure}[h]
\center{
\rotatebox{-90}{\resizebox{6cm}{!}{\includegraphics{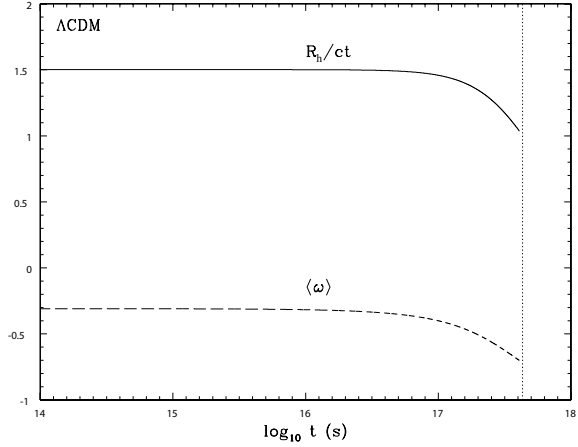}}}\\
\caption{Plot of the horizon radius $R_{\rm h}$ in units of $ct$, and
$\langle\omega\rangle$, the equation of state parameter $\omega\equiv p/\rho$
averaged over time from $t$ to $t_0$. The asymptotic value of $\langle\omega
\rangle$ (called $\langle\omega\rangle_\infty$ in the text) for $t\rightarrow 0$ 
is approximately $-0.31$. These results are from a calculation of the universe's 
expansion in a $\Lambda$CDM cosmology, with matter energy density $\rho_m(t_0)=
0.3\rho_c(t_0)$ and a cosmological constant $\rho_d=\rho_\Lambda=0.7\rho_c(t_0)$, 
with $\omega_d\equiv\omega_\Lambda=-1$.}
}
\end{figure}
\begin{figure}[h]
\center{
\rotatebox{-90}{\resizebox{6cm}{!}{\includegraphics{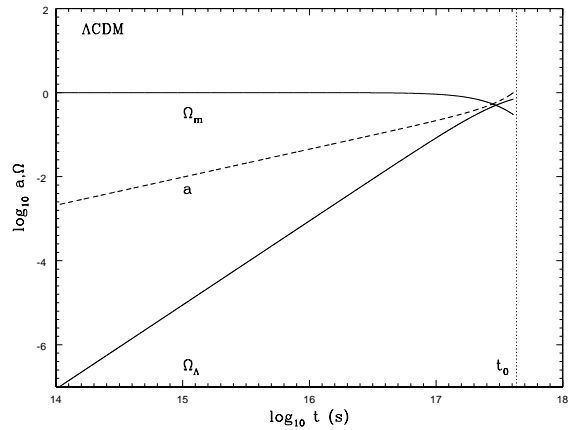}}}\\
\caption{Plot of the matter energy density parameter
$\Omega_m\equiv\rho_m(t)/\rho_c(t)$, the cosmological constant energy
density parameter $\Omega_\Lambda\equiv\rho_\Lambda(t)/\rho_c(t)$,
and the expansion factor $a(t)$, as functions of cosmic time $t$ for
the same $\Lambda$CDM cosmology as that shown in Fig.~1. Since the
energy density associated with $\Lambda$ is constant, $\Omega_m
\rightarrow 1$ as $t\rightarrow 0$. A notable (and well-known) peculiarity of
this cosmology is the co-called ``coincidence problem," so dubbed because
$\Omega_\Lambda$ is approximately equal to $\Omega_m$ only in the current
epoch.}
}
\end{figure}

But let us now see whether $\Lambda$CDM is also consistent with
our understanding of $R_{\rm h}$. Putting $\rho=\rho_m(t)+\rho_\Lambda$,
where $\rho_m$ is the time-dependent matter energy density, we may
integrate Eqn.~(8) for a $\Lambda$CDM cosmology, starting at the
present time $t_0$, and going backwards towards the era when radiation
dominated $\rho$ (somewhere around 100,000 years after the Big
Bang). Fig.~1 shows the run of $R_{\rm h}/ct$ as a function of time, along
with the time-averaged $\omega$ given in Eqn.~(10), to be distinguished
from the asymptotic value $\langle\omega\rangle_\infty$, which is the
equation-of-state parameter $\omega$ averaged over the entire universal
expansion, from $t=0$ to the present. The present epoch
is indicated by a vertical dotted line. The calculation begins at the
present time $t_0$, with the initial value $R_{\rm h}=(3/2)(1+
\langle\omega\rangle_\infty)ct_0$, where $\langle\omega\rangle_\infty$ 
is obtained by an iterative convergence of the solution 
to Eqn.~(8). In order to have $R_{\rm  h}=ct_0$ at the present time,
$\langle\omega\rangle_\infty$ must be ($\approx -0.31$). In $\Lambda$CDM,
the matter density increases towards the Big Bang, but $\rho_\Lambda$
is constant, so the impact of $\omega_d$ on the solution vanishes as
$t\rightarrow 0$ (see the dashed curve in Fig.~1). Thus, as expected,
$R_{\rm h}/ct\rightarrow 3/2$ early in the Universe's development. This
is the correct behavior within the framework of the
Friedman-Robertson-Walker cosmology.

What is rather striking about this result is that in $\Lambda$CDM,
$R_{\rm h}(t)$ approaches $ct$ only once in the entire history of the
Universe---and this is only because we have imposed this requirement
as an initial condition on our solution.  There are many peculiarities
in the standard model, some of which we will encounter shortly, but the
unrealistic coincidence that $R_{\rm h}$ should approach $ct_0$ only at the
present moment must certainly rank at---or near---the top of this list.

One may be tempted to think of this result as another manifestation of
the so-called ``coincidence problem" in $\Lambda$CDM cosmology, arising
from the peculiar near-simultaneous convergence of $\rho_m$ and
$\rho_\Lambda$ towards $\rho_c$ in the present epoch. But these are
definitely not the same phenomenon. Whether or not $\rho_m$ and
$\rho_\Lambda$ are similar, the requirement that $R_{\rm h}/ct\rightarrow 1$
only at $t_0$ implies a very special evolution of these quantities
as functions of cosmic time $t$ (see Fig.~2). This odd behavior casts
doubt on the viability of $\Lambda$CDM cosmology as the correct description
of the Universe.

\section{Dynamical Dark Energy}
Given the broad range of alternative theories of dark energy that
are still considered to be viable, it is beyond the scope of this paper
to exhaustively study all dynamical scenarios. Instead, we
shall focus on a class of solutions with particular importance to
cosmology---those in which the energy density of the scalar field mimics
the background fluid energy density. Cosmological models in this category
are known as ``scaling solutions," characterized by the relation
\begin{equation}
\frac{\rho_d(t)}{\rho_m(t)}=\frac{\rho_d(t_0)}{\rho_m(t_0)}\approx 2.33
\end{equation}
(some of the early papers on this topic include Refs.~18--25).

By far the simplest cosmology we can imagine in this class is that
for which $\omega=-1/3$, corresponding to $\omega_d\approx -1/2$ (within
the errors). This model is conceptually very attractive, but does not
appear to be fully consistent with Type Ia
supernova data, so either our interpretation of current observations
is wrong or the Universe just doesn't work this way. But it's worth
our while spending some time with it because it will help us understand
the models that follow it.
\begin{figure}[h]
\center{
\rotatebox{-90}{\resizebox{6cm}{!}{\includegraphics{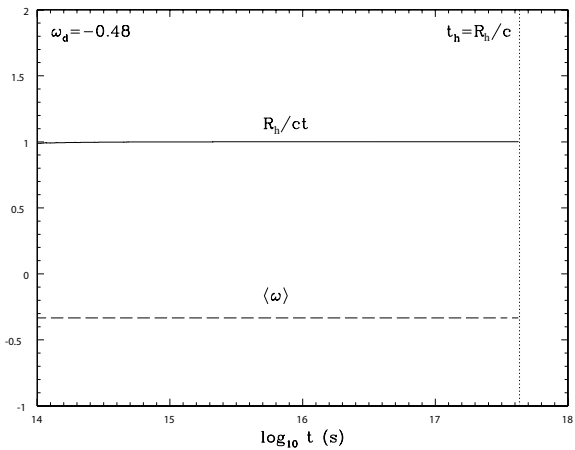}}}\\
\caption{Same as Fig.~1, except for a ``scaling solution"
in which $\rho_d\propto\rho_m$ and $\langle\omega\rangle=-1/3$. The
corresponding dark-energy equation-of-state parameter is $\omega_d\approx
-0.48$. This is sufficiently close to $-0.5$, that it may actually correspond
to this value within the errors associated with the measurement
of $\rho_m(t_0)/\rho_c(t_0)$ and $\rho_d(t_0)/\rho_c(t_0)$. Note that for this
cosmology, $R_{\rm h}/ct$ is always exactly one. A universe such as this would
do away with the otherwise inexplicable coincidence that $R_{\rm h}(t_0)=ct_0$
(since it has this value for all $t$), but as we shall see in Fig.~5, it does
not appear to be entirely consistent with Type Ia supernova data.}
}
\end{figure}
\begin{figure}[h]
\center{
\rotatebox{-90}{\resizebox{6cm}{!}{\includegraphics{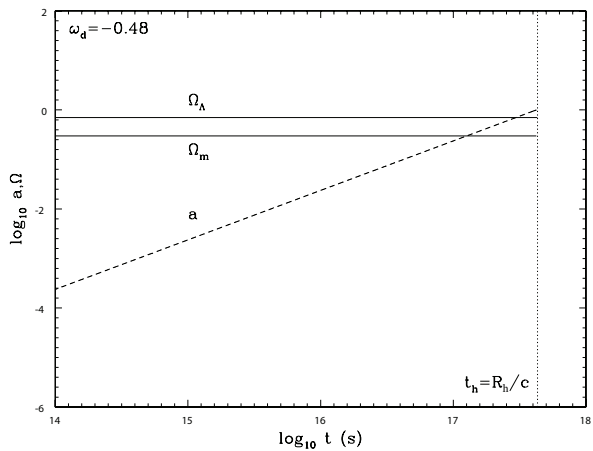}}}\\
\caption{Same as Fig.~2, except for a ``scaling solution"
in which $\rho_d\propto\rho_m$ and $\langle\omega\rangle=-1/3$. The
corresponding dark-energy equation-of-state parameter is $\omega_d\approx
-0.48$.}
}
\end{figure}

To begin with, we see immediately from Eqn.~(8) (and illustrated in
Figs.~3 and 4) that when $\omega=-1/3$, we have $R_{\rm h}(t)=ct$. Thus, for
a scaling solution satisfying Eqn.~(12), this value of $\omega$ solves three 
of the major problems in standard cosmology: first, it explains why $R_{\rm h}(t_0)$ 
should be equal to $ct_0$ (because these quantities are always equal). Second, it 
removes the inexplicable coincidence that $\rho_d$ and $\rho_m$ should be 
comparable to each other only in the present epoch (since they are always 
comparable to each other). Third, it does away with the so-called flatness 
problem. To see this, let us return momentarily to Eqn.~(1) and rewrite it
as follows:
\begin{equation}
H^2=\left(\frac{c}{R_{\rm h}}\right)^2\left(1-\frac{kR_{\rm h}^2}{a^2}\right)\;.
\end{equation}
Whether or not the Universe is asymptotically flat hinges on the
behavior of $R_{\rm h}/a$ as $t\rightarrow\infty$. But from the definition
of $R_{\rm h}$ (Eqn.~7), we infer that
\begin{equation}
\frac{d}{dt}\ln R_{\rm h}=\frac{3}{2}(1+\omega)\frac{d}{dt}\ln a\;.
\end{equation}
Thus, if $\omega=-1/3$,
\begin{equation}
\frac{d}{dt}\ln R_{\rm h}=\frac{d}{dt}\ln a\;,
\end{equation}
and so
\begin{equation}
H={\rm constant}\times \left(\frac{c}{R_{\rm h}}\right)\;,
\end{equation}
which is the equation for a flat universe. (We caution, however,
that in deriving this result, we have used Eqns.~8 and 11, which
implicitly assume a flat universe, together with Eqn.~13 which
is more general. If a scaling solution proves to be the likely
representation of dark energy, then a demonstration of flatness
would need to be shown more rigorously.) We also learn from Eqn.~(2) 
that in this universe, $\ddot{a}=0$. The Universe is coasting, but
not because it is empty, as in the Milne cosmology,$^{27}$
but rather because the change in pressure as it expands is just right to
balance the change in its energy density.

All told, these are quite impressive accomplishments for such a simple
model, and yet, it doesn't appear to be fully consistent with Type Ia
supernova data. It is quite straightforward to demonstrate this with
the evolutionary profiles shown in Figs.~3 and 4.  The comoving
coordinate distance from some time $t$ in the past to the present is
$\Delta r=\int_t^{t_0} c\,dt/a(t)$. With $k=0$, the luminosity
distance $d_L$ is $(1+z)\Delta r$, where the redshift $z$ is given by
$(1+z)=1/a$, in terms of the expansion factor $a(t)$ plotted in Fig.~4.

The data in Fig.~5 are taken from the ``gold" sample,$^{28}$
with coverage in redshift from $0$ to $\sim$1.8.
The distance modulus is $5\log d_L(z)+\Delta$, where $\Delta\approx
25$. The dashed curve in this plot represents the fit based on the
scaling solution shown in Figs.~3 and 4, with a Hubble
constant $H_0=71$ km s$^{-1}$ Mpc$^{-1}$ ($\Delta$ is used as
a free optimization parameter in each of Figs.~5 and 8). The
``best" match corresponds to an unacceptable reduced $\chi^2$ of
$1.11$ with $180-1=179$ d.o.f.

Interestingly, if we were to find a slight systematic
error in the distance modulus for the events at $z>1$, which for some
reason has led to a fractional over-estimation in their distance (or,
conversely, a systematic error that has lead to an under-estimation
of the distance modulus for the nearby explosions), the fit would
improve significantly. So our tentative conclusion right now must
be that, although an elegant scaling solution with $\omega=-1/3$
provides a much better explanation than $\Lambda$CDM for the observed
coincidence $R_{\rm h}\approx ct_0$, it is nonetheless still not fully
consistent with the supernova data.

\begin{figure}[h]
\center{
\rotatebox{-90}{\resizebox{6cm}{!}{\includegraphics{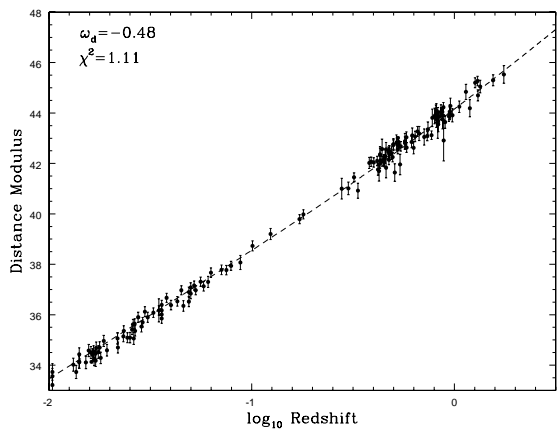}}}\\
\caption{Plot of the observed distance modulus versus redshift for
well-measured distance Type Ia supernovae.$^{28}$
The dashed curve shows the theoretical distribution of magnitude versus
redshift for the ``scaling solution" in which $\rho_d\propto\rho_m$ and
$\langle\omega\rangle=-1/3$ (see Figs.~3 and 4). The corresponding
dark-energy equation-of-state parameter is $\omega_d\approx
-0.48$. The best fit corresponds to a reduced $\chi^2\approx 1.11$ for
$180-1=179$ d.o.f., which is not an adequate representation of
the Type Ia supernova data.}
}
\end{figure}

Fortunately, many of the attractive features of an $\omega=-1/3$
cosmology are preserved in the case where $\omega_d=-2/3$,
corresponding to a scaling solution with $\omega\approx-1/2$.
This model fits the supernova data quite strikingly, but it comes
at an additional cost---we would have to accept the fact that
the universe is somewhat older (by a few billion years) than we
now believe. Actually, this situation is unavoidable for any cosmology
with $\omega<-1/3$ because of the relation between $R_{\rm h}$ and
$ct_0$ in Eqn.~(11). This conclusion may seem incompatible
with the WMAP and HST data, but is actually fully consistent with them,
though our \emph{interpretation} of the currently inferred ``age"
of 13.7 billion years would need to be revised along the following
lines.

Light reaching us from beyond the cosmic horizon (at radius $R_{\rm h}$)
is infinitely redshifted, so only phenomena occurring within a time
$t_h\equiv R_{\rm h}/c$ of the present can produce a measurable signal in
our instruments. Of course, there is no \emph{a priori} reason why
the horizon time $t_h$ must coincide with the time $t_0$ elapsed since
the Big Bang. We can say with certainty that $t_0$ cannot be
smaller than $t_h$, for otherwise $R_{\rm h}>ct_0$, which is inconsistent
with the observations. However, a situation with $t_0>t_h$ simply
means that the Universe has been expanding longer than the time
it has taken the observed CMB to reach us from $R_{\rm h}$. As noted
earlier, this situation cannot be avoided for any scenario in
which $\omega<-1/3$. The case $\omega=-1/3$ is special because
then $R_{\rm h}=ct_0$.

\begin{figure}[h]
\center{
\rotatebox{-90}{\resizebox{6cm}{!}{\includegraphics{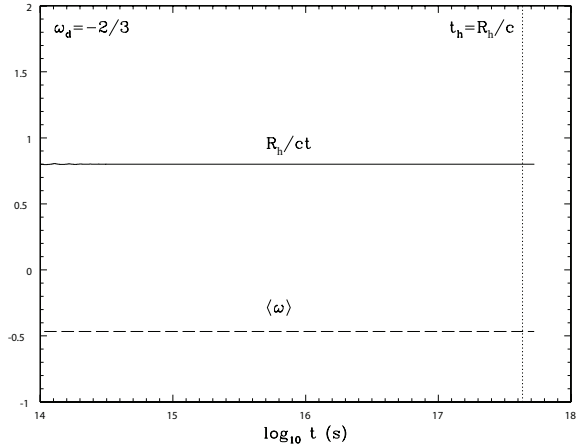}}}\\
\caption{Same as Fig.~1, except for a ``scaling solution"
in which $\rho_d\propto\rho_m$ and $\omega_d=-2/3$. The time-averaged
equation-of-state parameter is $\langle\omega\rangle\approx -0.47$.
Thus, $R_{\rm h}\not= ct_0$. Instead, $t_0\approx 16.9$ billion years,
approximately $3.4$ billion years longer than the ``horizon" time,
$t_h\equiv R_{\rm h}/c$ ($\approx 13.5$ billion years). This type of
universe is not subject to the ``coincidence" problem since $R_{\rm h}/ct$
is constant. It provides the best fit to the Type Ia supernova data
(see Fig.~8).}
}
\end{figure}

We see in Fig.~6 that $R_{\rm h}/ct$ is constant (say when $\omega_d=-2/3$),
but at a value $(3/2)(1+\langle\omega\rangle)$, where the time-averaged
$\omega$ is now $\approx -0.47$. Thus, if $R_{\rm h}$ is 13.5 billion light-years,
$t_0$ must be approximately 16.9 billion years. This is simply
a manifestation of the fact that anything that happened between the
Big Bang and a horizon time $t_h$ ago would have produced infinitely
redshifted signals at the present time, and is no longer observable.
Thus, we can see only as far back as $t_h$,
and if $\omega<-1/3$, we must therefore re-interpret the presently
inferred ``age" as the horizon time $t_h$, not the Universe's true
age $t_0$. In Fig.~6, the distinction between $t_h$ and $t_0$ is
indicated primarily through the termination points of the $R_{\rm h}$ and
$\langle\omega\rangle$ curves, which extend past the vertical dotted
line at $t=t_h$.

Ironically, this unexpected result has several important consequences,
such as offering an explanation for the early appearance of supermassive
black holes$^{29}$ (at a redshift $>6$), and the glaring
deficit of dwarf halos in the local group.$^{30}$
Both of these long-standing problems in cosmology would be resolved if
the Universe were older. Supermassive black holes would have had much
more time ($4-5$ billion years) to form than current thinking allows
(i.e., only $\sim 800$ million years), and dwarf halos would
correspondingly have had more time to merge hierarchically, depleting
the lower mass end of the distribution.

\begin{figure}[h]
\center{
\rotatebox{-90}{\resizebox{6cm}{!}{\includegraphics{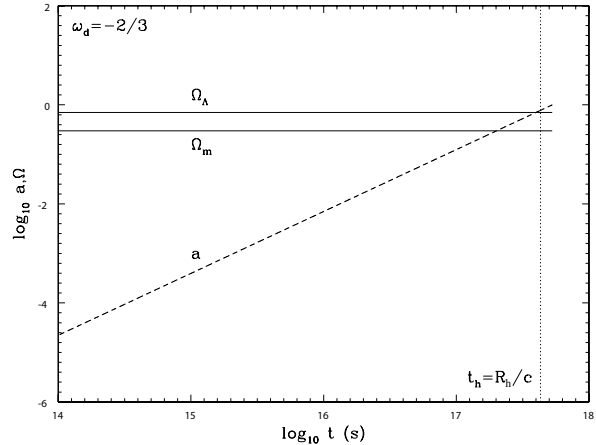}}}\\
\caption{Same as Fig.~2, except for a ``scaling solution"
in which $\rho_d\propto\rho_m$ and $\omega_d=-2/3$.}
}
\end{figure}

The matter and dark energy densities corresponding to the $\omega_d=-2/3$
scaling solution are shown as functions of cosmic time in Fig.~7, along
with the evolution of the scale factor $a(t)$. Here too, the ``coincidence
problem" does not exist, and the flatness problem is resolved since
$R_{\rm h}/a\rightarrow 0$ as a result of Eqn.~(14), so that 
$kR_{\rm h}^2/a^2\rightarrow 0$ as $t\rightarrow\infty$ in Eqn.~(13). 
Thus, the constant in Eqn.~(16) should be $\approx 1$ at late times,
regardless of the value of $k$. Very importantly, this model fits the
Type Ia supernova data very well, as shown in Fig.~8. The best fit
corresponds to $\Delta=25.26$, with a reduced $\chi^2=1.001$ for
$180-1=179$ d.o.f.

\section{Concluding Remarks}

Our main goal in this study has been to examine what we can learn
about the nature of dark energy from a consideration of the cosmic
horizon $R_{\rm h}$ and its evolution with cosmic time. A principal
outcome of this work is the realization that the so-called
``coincidence" problem in the standard model is actually more
severe than previously thought. We have found that in a
$\Lambda$CDM universe, $R_{\rm h}\rightarrow ct_0$ only once, and
according to the observations, this must be happening right now.
The unlikelihood of this occurrence is an indication that dark
energy almost certainly is not due to a cosmological constant.
Other issues that have already been discussed extensively in
the literature, such as the fact that the vacuum energy in
quantum theory should greatly exceed the required value of
$\Lambda$, only make this argument even more compelling. Of
course, this rejection of the cosmological-constant hypothesis
then intensifies interest into the question of why we don't
see any vacuum energy at all, but this is beyond the scope
of the present paper.

\begin{figure}[h]
\center{
\rotatebox{-90}{\resizebox{6cm}{!}{\includegraphics{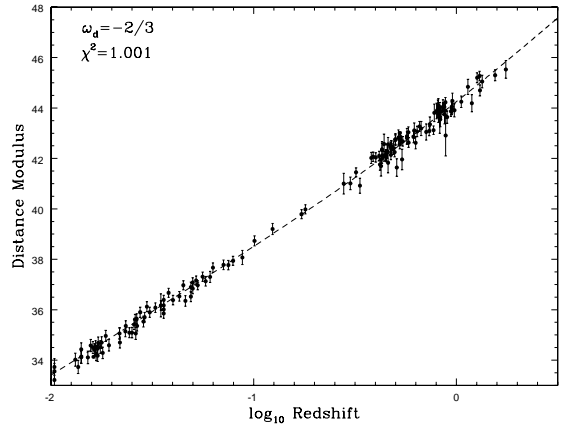}}}\\
\caption{Same as Figure~5, except for a ``scaling solution"
in which $\rho_d\propto\rho_m$ and $\omega_d=-2/3$. This type of universe
provides the best fit to the Type Ia supernova data,$^{28}$
with a reduced $\chi^2\approx 1.001$ for $180-1=179$ d.o.f. The best fit
corresponds to $\Delta=25.26$.}
}
\end{figure}

Many alternatives to a cosmological constant have been
proposed over the past decade but, for the sake of simplicity,
we have chosen in this paper to focus our attention on scaling
solutions. The existence of such cosmologies has been discussed
extensively in the literature, within the context of standard
General Relativity, braneworlds (Randall-Sundrum and Gauss-Bonnett),
and Cardassian scenarios, among others.$^{31,32,33}$

Our study has shown that scaling solutions fit the Type Ia supernova
data at least as well as the basic $\Lambda$CDM cosmology, but they
go farther in simultaneously solving several conundrums with the
standard model. As long as the time-averaged
value of $\omega$ is less than $-1/3$, they eliminate both the
coincidence and flatness problems, possibly even obviating the
need for a period of rapid inflation in the early universe.$^{34,35}$

But most importantly, as far as this study is concerned, scaling
solutions account very well for the observed fact that $R_{\rm h}\approx
ct_0$. If $\langle\omega\rangle=-1/3$ exactly, then $R_{\rm h}(t)=ct$
for all cosmic time, and therefore the fact that we see this
condition in the present Universe is no coincidence at all.
On the other hand, if $\langle\omega\rangle<-1/3$, scaling
solutions fit the Type Ia supernova data even better, but
then we have to accept the conclusion that the Universe is
older than the horizon time $t_h\equiv R_{\rm h}/c$. According to
our calculations, which produce a best fit to the supernova
data for $\langle\omega\rangle=-0.47$ (corresponding to a
dark energy equation of state with $\omega_d=-2/3$), the age
of the Universe should then be $t_0\approx 16.9$ billion years.
This may be surprising at first, but the fact of the matter
is that such an age actually solves other major problems
in cosmology, including the (too) early appearance of
supermassive black holes, and the glaring deficit of
dwarf halos in the local group of galaxies.

When thinking about a dynamical dark energy, it is worth recalling
that scalar fields arise frequently in particle physics, including
string theory, and any of these may be appropriate candidates for
this mysterious new component of $\rho$. Actually, though we have
restricted our discussion to equations of state with $\omega_d\ge -1$,
it may even turn out that a dark energy with $\omega_d<-1$ is
providing the Universe's acceleration. Such a field is usually
referred to as a Phantom or a ghost. The simplest explanation
for this form of dark energy would be a scalar field with a
negative kinetic energy.$^{36}$ However, Phantom
fields are plagued by severe quantum instabilities, since
their energy density is unbounded from below and the vacuum
may acquire normal, positive energy fields.$^{37}$
We have therefore not included theories with $\omega_d<-1$
in our analysis here, though a further consideration of their
viability may be warranted as the data continue to improve.

On the observational front, the prospects for confirming or
rejecting some of the ideas presented in this paper look
very promising indeed. An eagerly anticipated mission, SNAP,$^{38}$
will constrain the nature of dark energy
in two ways. First, it will observe deeper Type Ia supernovae.
Second, it will attempt to use weak gravitational lensing
to probe foreground mass structures. If selected, SNAP should
be launched by 2020. The Planck CMB satellite, an already funded
mission, probably won't have the sensitivity to measure
any evolution in $\omega_d$, but it may be able to tell us
whether or not $\omega_d=-1$. 

Finally, we may be on the verge of uncovering a class
of sources other than Type Ia supernovae to use for
dark-energy exploration. Type Ia supernovae have greatly
enhanced our ability to study the Universe's expansion out
to a redshift $\sim 2$. But this new class of sources may
possibly extend this range to values as high as 5--10.
According to Ref.~39, Gamma Ray Bursts
(GRBs) have the potential to detect dark energy with a
reasonable significance, particularly if there was an
appreciable amount of it at early times, as suggested by
scaling solutions. It is still too early to tell if GRBs
are good standard candles, but since differences between
$\Lambda$CDM and dynamical dark energy scenarios are more
pronounced at early times (see Figs.~5 and 8),
GRBs may in the long run turn out to be even more important
than Type Ia supernovae in helping us learn about the true
nature of this unexpected ``third" form of energy.

\section*{Acknowledgments}

This research was partially supported by NSF grant 0402502 at the
University of Arizona. Many inspirational discussions with Roy Kerr
are greatly appreciated. I also thank the anonymous referee for his
helpful comments. Part of this work was carried out at Melbourne 
University and at the Center for Particle Astrophysics and Cosmology 
in Paris. 


\end{document}